# IMPROVING UPON THE EFFECTIVE SAMPLE SIZE BASED ON GODAMBE INFORMATION FOR BLOCK LIKELIHOOD INFERENCE

Rahul Mukerjee
Indian Institute of Management Calcutta
Joka, Diamond Harbour Road
Kolkata 700 104, India
E-mail: rmuk0902@gmail.com

*Abstract*: We consider the effective sample size, based on Godambe information, for block likelihood inference which is an attractive and computationally feasible alternative to full likelihood inference for large correlated datasets. With reference to a Gaussian random field having a constant mean, we explore how the choice of blocks impacts this effective sample size. It is seen that spreading out the spatial points within each block, instead of keeping them close together, can lead to considerable gains while retaining computational simplicity. Analytical results in this direction are obtained under the AR(1) model. The insights so found facilitate the study of other models, including correlation models on a plane, where closed form expressions are intractable.

**1. Introduction**

The notion of effective sample size (ESS) dates back to Kish (1965, Chapter 8) who defined it, for a given sampling design, as the sample size required by simple random sampling to achieve the same variance of a parameter estimator as in the given design. In the context of identically distributed but correlated observations, it can be viewed as the number of observations that would carry the same information on a parameter of interest, had the observations been independent. Thus ESS plays a crucial role in assessing the information content of the data, eliminating the duplicated information due to correlation. As a result, an understanding of ESS becomes important whenever the data are likely to show an appreciable amount of correlation as happens in diverse areas such as time series and spatial data analysis (Cressie, 1993, Chapters 1, 4), repeated measurements (Faes et al., 2009), Bayesian model selection (Berger et al., 2014), importance sampling and MCMC (Martino et al., 2017; Chatterjee and Diaconis, 2018), to name a few.

The present work is inspired by a recent paper by Acosta et al. (2021) who proposed and investigated an ingenious and useful formulation of ESS, via Godambe information, for block likelihood inference. This was done in the framework of a Gaussian random field having a constant mean. Earlier, for such a random field, Vallejos and Osorio (2014) studied ESS on the basis of Fisher information for full likelihood inference. However, as noted by Acosta et al. (2021), full likelihood inference becomes infeasible for large correlated datasets because of the challenge in inverting the correlation matrix. They suggested block likelihood inference (Caragea and Smith, 2007; Varin et al., 2011) as a statistically efficient and viable alternative that requires inversion of smaller correlation matrices and hence significantly reduces the computational burden. Interestingly, they showed that the resulting Godambe information-based ESS underestimates but, under a wide variety of parametric correlation structures, well approximates the Fisher information-based ESS for full likelihood inference. See their paper for more details on block likelihood inference as well as a concise but informative review of ESS in general, with further references.

For ease in presentation, hereafter, the Godambe information-based formulation of ESS in Acosta et al. (2021) for block likelihood inference is called $ESS_B$, while the Fisher information-based ESS in Vallejos and Osorio (2014) for full likelihood inference is referred to simply as ESS.

In their approach to block likelihood inference, Acosta et al. (2021), kept the spatial points within each block as close together as possible; see their Sections 4 and 5. At the same time, they remarked in their Section 3 that $ESS_B$ can strongly depend on the block partitioning. From this perspective, we explore how the choice of blocks impacts $ESS_B$. It is found that spreading out the spatial points within each block, instead of keeping them close together, can lead to considerable gains in $ESS_B$, thus entailing still better approximations to ESS. Analytical results in this direction are obtained in Section 3 under the AR(1) model, after presenting the preliminaries in Section 2. The ideas emerging from these analytical results facilitate the study of other models, such as one-dimensional correlation models in Section 4, and correlation models on a plane in Section 5, where closed form expressions are intractable. Finally, the paper ends in Section 6 with some concluding remarks.

**2. Preliminaries**

As in Vallejos and Osorio (2014) and Acosta et al. (2021), consider a spatial random field $\{X(s), s \in \mathcal{R}^p\}$, and let $X_i = X(s_i)$, $i = 1,\ldots, n$, be a realization at $n$ spatial points $s_1,\ldots, s_n$. Write $X = (X_1,\ldots, X_n)^T$, where the superscript T stands for transpose. Suppose the random field is Gaussian with a constant mean $\mu \in \mathcal{R}$, and a constant variance $\sigma^2$ ($> 0$). Then $X$ is multivariate normal with $E(X) = \mu 1_n$ and $cov(X) = \sigma^2 R$, where $R = R(\theta)$ is the $n \times n$ correlation matrix of $X$, with $\theta$ as a possibly vector-valued parameter, and $1_a$ is the $a \times 1$ vector of ones for any positive integer $a$. In what follows, $R$ is supposed to be positive definite.

In the above setup, based on a normalized version of Fisher information about $\mu$ for full likelihood inference, Vallejos and Osorio (2014) defined ESS as

$$\text{ESS} = 1_n^T R^{-1} 1_n. \qquad (1)$$

As mentioned in the introduction, however, full likelihood inference becomes infeasible for large correlated datasets. From this viewpoint, Acosta et al. (2021) considered a block likelihood approach based on a partition $\{B_1,\ldots, B_m\}$ of $\{1,\ldots, n\}$ and the resulting product of likelihoods of the data blocks $X_{(u)} = \{X_i : i \in B_u\}$, $u = 1,\ldots, m$. This can be viewed as a misspecified likelihood which pretends independence among the data blocks $X_{(1)},\ldots, X_{(m)}$. Each subset $B_u$, hereafter called a block, signifies a block of the data, and hence the partition is also called a blocking. For $u, v = 1,\ldots, m$, let $R_{uv}$ be the $b_{(u)} \times b_{(v)}$ cross-correlation matrix of $X_{(u)}$ and $X_{(v)}$, where $b_{(u)}$ is the cardinality of $B_u$ and $b_{(1)} +\ldots+ b_{(m)} = n$. Note that each $R_{uv}$ is a submatrix of $R$ and any $R_{uu}$ is the correlation matrix of $X_{(u)}$. Based on a normalized version of the Godambe information about $\mu$ for block likelihood inference, Acosta et al. (2021) defined the effective sample size for the blocking $\{B_1,\ldots, B_m\}$ as

$$\text{ESS}_B = \frac{(\sum_{u=1}^m \lambda_{uu})^2}{\sum_{u=1}^m \sum_{v=1}^m \lambda_{uv}}, \tag{2}$$

where

$$\lambda_{uv} = 1_{b_{(u)}}^T R_{uu}^{-1} R_{uv} R_{vv}^{-1} 1_{b_{(v)}}, \qquad u, v = 1,\ldots, m. \tag{3}$$

They observed that if $m = 1$, then block likelihood reduces to full likelihood and

$$\text{ESS}_B = \text{ESS} = 1_n^T R^{-1} 1_n, \tag{4}$$

while, at the other extreme, if $m = n$, then (2) simplifies to

$$\text{ESS}_B = n^2 / (1_n^T R 1_n). \tag{5}$$

Acosta et al. (2021) also noted that more generally, $\text{ESS}_B \leq \text{ESS}$ for every blocking; cf. Godambe and Kale (1991, pp. 3-20). Hence it makes sense to define the efficiency of any such blocking as the ratio

$$\text{Eff}_B = \text{ESS}_B / \text{ESS}. \tag{6}$$

Naturally, for any given number of blocks $m$ and block sizes $b_{(1)},\ldots, b_{(m)}$, a blocking with a higher $\text{ESS}_B$, and hence an $\text{Eff}_B$ which is closer to 1, is preferred.

Let $R = (r_{ij})$. In the next section, we focus attention on the AR(1) model, where $r_{ij} = \rho^{|i-j|}$ ($i, j = 1,\ldots, n$) and $0 \leq \rho < 1$. This is a one-dimensional stationary correlation model where $r_{ij}$ depends on $i$ and $j$ only through $|i - j|$, i.e.,

$$r_{ij} = r_{|i-j|}, \quad i, j = 1,\ldots, n, \tag{7}$$

for scalars $r_0, r_1,\ldots, r_{n-1}$, with $r_0 = 1$. The AR(1) model allows us to obtain analytical results on $\text{ESS}_B$ which, in turn, facilitate an understanding of other stationary correlation models as in (7). Like AR(1), these models are natural candidates for consideration, for instance, when the $n$ spatial points are equispaced over $\mathcal{R}$. Moreover, as seen in Section 5, the insights gained through the AR(1) model apply to two-dimensional correlation models as well. An extension of the AR(1) model to the case of non-equispaced spatial points appears in Subsection 4.2.

For any $h$ ($\geq 0$) and any integer $a$ ($\geq 2$), write

$$y_a(h) = \{(1-h)1_a + h\psi_a\}/(1+h), \tag{8}$$

where $\psi_a$ is the $a \times 1$ vector with 1 at the first and last positions and zeros elsewhere.

**Proposition 1**. *Under the AR(1) model,*

(a) $R^{-1}1_n = y_n(\rho)$, (b) ESS = $\{n(1-\rho) + 2\rho\}/(1+\rho)$.

*Proof.* Under the AR(1) model, for $i = 1,\ldots, n$, the elements in the $i$th row of $R$ sum to

$$R[i] = \Sigma_{j=1}^{n} \rho^{|i-j|} = \Sigma_{j=1}^{i-1} \rho^{i-j} + \Sigma_{j=i}^{n} \rho^{j-i} = (1 + \rho - \rho^i - \rho^{n-i+1})/(1-\rho), \tag{9}$$

so that, by (8), the $i$th element of $Ry_n(\rho)$ equals

$$\{(1-\rho)R[i] + \rho(\rho^{i-1} + \rho^{n-i})\}/(1+\rho) = 1,$$

and (a) follows. Now, (b) is immediate from (1) and (8). □

Part (b) of Proposition 1 is available in Vallejos and Osorio (2014), while part (a) will be seen to have implications later in a proof in the appendix. .

## 3. Row- and column-wise blockings

### 3.1. Two kinds of blocking

In this paper, we shall be primarily concerned with blocks of equal size, though the case of unequal block size will also be discussed later in Subsection 4.3. Let $n = mb$, where $m$ and $b$ are positive integers, and suppose it is intended to have $m$ blocks each of size $b$.

In the framework of a one-dimensional correlation model such as the AR(1) model, we consider two kinds of blocking, namely, row-wise (RW) blocking and column-wise (CW) blocking. The RW blocking partitions $\{1,\ldots, n\}$ into the $m$ blocks, each of size $b$, as given by

$$B_u^{\text{row}} = \{(u-1)b + j : j = 1,\ldots, b\}, \quad u = 1,\ldots, m. \tag{10}$$

We call it RW as these blocks correspond to the rows of an arrangement of $1,\ldots, n$ in the natural order as an $m \times b$ array, e.g., if $m = 3$ and $b = 5$, then they are given by the rows of

$$\begin{array}{ccccc} 1 & 2 & 3 & 4 & 5 \\ 6 & 7 & 8 & 9 & 10 \\ 11 & 12 & 13 & 14 & 15 \end{array}.$$

The CW blocking, on the other hand, partitions $\{1,\ldots, n\}$ into the $m$ blocks

$$B_u^{\text{col}} = \{u + (j-1)m : j = 1,\ldots, b\}, \quad u = 1,\ldots, m, \tag{11}$$

each of size $b$. It is called CW as these blocks correspond to the columns of an arrangement of $1,\ldots, n$ in the natural order as a $b \times m$ array, e.g., if $m = 3$ and $b = 5$, then they are given by the columns of

$$\begin{array}{ccc} 1 & 2 & 3 \\ 4 & 5 & 6 \\ 7 & 8 & 9 \\ 10 & 11 & 12 \\ 13 & 14 & 15 \end{array}.$$

For the AR(1) model, Acosta et al. (2021, Subsection 5.2) considered RW blocking which is, indeed, more intuitive than CW blocking; e.g., with $m = 3$ and $b = 5$, the blocking $\{1,2,3,4,5\}$, $\{6,7,8,9,10\}$, $\{11,12,13,14,15\}$ certainly looks more natural than $\{1,4,7,10,13\}$, $\{2,5,8,11,14\}$, $\{3,6,9,12,15\}$, where the points in each block are much more spread out. Our analytical and computational results, however, show that CW blocking can entail considerable improvement over RW blocking with regard to $ESS_B$ under the AR(1) model and beyond. Moreover, this improvement in $ESS_B$ comes without sacrificing the features of RW blocking that simplify computation in these models; see Remark 1 below. It is also seen in Section 5 that there is scope for a gainful extension of the idea of CW blocking to two- or potentially multi-dimensional correlation models.

We denote the $ESS_B$ for RW and CW blockings by $ESS_{row}$ and $ESS_{col}$, respectively. Also, following (6), the efficiencies of these blockings are defined as

$$\text{Eff}_{row} = ESS_{row}/ESS, \qquad \text{Eff}_{col} = ESS_{col}/ESS. \tag{12}$$

If $(b, m) = (n, 1)$ or $(1, n)$, then these two blockings become identical and they both satisfy (4) or (5), respectively. Interestingly, as the next result shows, both these blockings satisfy (5) for $b = 2$ as well, under any stationary correlation model of the form (7).

**Proposition 2**. *If $b = 2$, then under any stationary correlation model,*

$$ESS_{row} = ESS_{col} = n^2/(1_n^T R 1_n).$$

*Proof.* Let (7) hold. By (10) and (11), then for both RW and CW blockings, $R_{11} = \ldots = R_{mm}$. Hence if $b = 2$, then for each of these blockings, the vectors $R_{11}^{-1} 1_b, \ldots, R_{mm}^{-1} 1_b$ equal the same scalar multiple of $1_b$. The result now follows from (2) and (3), on simplification. □

**Remark 1**. In the spirit of Proposition 2, even for general $b$, under any stationary correlation model, CW blocking enjoys the same simplifying features as RW blocking. For both these blockings, by (10) and (11), under such a model, any submatrix $R_{uv}$ of $R$ depends on $u$ and $v$ only through $u - v$. Therefore, due to the symmetry of $R$, it follows from (2) and (3) that $ESS_{row}$ and $ESS_{col}$ can be computed by obtaining the respective $\lambda_{11}, \ldots, \lambda_{1m}$ alone. □

*3.2. The AR(1) model*

We now compare RW and CW blockings under the AR(1) model. For this model, by (9),

$$1_n^T R 1_n = \Sigma_{i=1}^n R[i] = \{n(1-\rho^2) - 2\rho(1-\rho^n)\}/(1-\rho)^2, \tag{13}$$

Hence, from (5) and Proposition 2, the following is evident.

**Proposition 3**. *If $b = 1$ or $2$, then under the AR(1) model,*

$$ESS_{row} = ESS_{col} = \frac{n^2(1-\rho)^2}{n(1-\rho^2) - 2\rho(1-\rho^n)}.$$

In view of Proposition 3, under the AR(1) model, one needs to compare $ESS_{row}$ and $ESS_{col}$ for $3 \leq b < n$. Theorem 1, proved in the appendix, presents useful analytical results for this purpose.

**Theorem 1**. *Under the AR(1) model, let $n = mb$, where $3 \leq b < n$. Then the $ESS_B$ for RW and CW blockings are given, respectively, by*

$$ESS_{row} = \frac{\{n(1-\rho) + 2m\rho\}^2}{(1+\rho)\left[n(1-\rho) + 2m\rho + \frac{2\rho(1+\rho)}{1-\rho^b}\left(m - \frac{1-\rho^n}{1-\rho^b}\right)\right]},$$

$$ESS_{col} = \frac{\{n(1-\rho^m) + 2m\rho^m\}^2 (1-\rho)^2}{(1-\rho^2)\{(n-2m)(1-\rho^m)^2 + 2m\} - 2\rho(1-\rho^{2m})}.$$

**Remark 2**. Theorem 1 remains valid for $b = n$ and $b = 2$ as well, with the expressions for $ESS_{row}$ and $ESS_{col}$ obtained there reducing, respectively, to their counterparts in (4) and Proposition 3. For $b = 1$, however, only the expression for $ESS_{row}$ in Theorem 1, and not that of $ESS_{col}$ in Theorem 1, matches its counterpart in Proposition 3. This is because the proof of Theorem 1 for $ESS_{col}$, unlike that for $ESS_{row}$, critically involves $\psi_b$ which is defined only for $b \geq 2$. Also, as expected, ESS, $ESS_{row}$ and $ESS_{col}$ in Proposition 1(b), Proposition 3 and Theorem 1 equal $n$ when $\rho = 0$. Furthermore, applying L'Hospital's rule, one can verify that each of these quantities tends to 1 as $\rho \to 1-$. □

The explicit results in Theorem 1 immediately allow comparison of $ESS_{row}$ and $ESS_{col}$ under the AR(1) model, for any $n$, $b$, $m$ and $\rho$. Equivalently, one can as well compare the corresponding efficiencies $Eff_{row}$ and $Eff_{col}$, relative to ESS, using (12) and Proposition 1(b). Our findings, which show clear gains via CW blocking under the AR(1) model, are summarized below:

(a) For every $n \leq 100000$, $3 \leq b \leq n/2$, i.e., $m \geq 2$, and every $\rho$ in $\{0.001, 0.002,\ldots, 0.999\}$, $ESS_{col}$ exceeds $ESS_{row}$, and hence $Eff_{col}$ exceeds $Eff_{row}$, their difference being more conspicuous for relatively higher values of $\rho$, e.g., if $(b, m) = (30, 30)$, then for $\rho = 0.6, 0.7, 0.8$ and $0.9$, the pair $(Eff_{row}, Eff_{col})$ equals $(0.961, 0.999)$, $(0.941, 0.998)$, $(0.919, 0.996)$ and $(0.913, 0.992)$, i.e., $Eff_{col}$ exceeds $Eff_{row}$, or equivalently, $ESS_{col}$ exceeds $ESS_{row}$ by 3.95%, 6.06%, 8.38% and 8.65%, respectively.

(b) Given $n$, $b$ and $m$, let $minEff_{row}$ and $minEff_{col}$ be the smallest possible $Eff_{row}$ and $Eff_{col}$, respectively, over $\rho$ in $\{0.001, 0.002,\ldots, 0.999\}$. These reflect the worst-scenario performances of RW and CW blockings and are of interest as $\rho$ is unknown in practice. For $n \leq 100000$, $minEff_{col}$ exceeds 0.9, 0.95 and 0.98 whenever $b \geq 3$, 9 and 26, respectively. On the other hand, $minEff_{row}$ can be considerably smaller even for relatively large $b$, e.g., for $(b, m) = (20, 10)$, $(30, 30)$ and $(60, 40)$, the pair $(minEff_{row}, minEff_{col})$ equals $(0.879, 0.980)$, $(0.878, 0.984)$ and $(0.878, 0.991)$, respectively.

(c) For $n \leq 100000$, the difference ESS – $\text{ESS}_{\text{col}}$ never exceeds 0.497, irrespective of $b$ ($\geq 3$), $m$ and $\rho$ in $\{0.001, 0.002,\ldots, 0.999\}$. Remarkably, this closeness of $\text{ESS}_{\text{col}}$ to ESS holds even without normalization through division by $n$. In contrast, the difference ESS – $\text{ESS}_{\text{row}}$ tends to be much larger, e.g., for $(b, m) = (20, 10)$, $(30, 30)$ and $(60, 40)$, the pair ($\text{maxDiff}_{\text{row}}$, $\text{maxDiff}_{\text{col}}$) equals (2.404, 0.314), (9.366, 0.415) and (17.261, 0.433), respectively, where $\text{maxDiff}_{\text{row}}$ is the largest possible value of ESS – $\text{ESS}_{\text{row}}$ over $\rho$ in $\{0.001, 0.002,\ldots, 0.999\}$ and $\text{maxDiff}_{\text{col}}$ is similarly defined.

(d) The gains via CW blocking persist even for $n > 100000$. For instance, if $(b, m) = (100, 5000)$, then ($\text{minEff}_{\text{row}}$, $\text{minEff}_{\text{col}}$) = (0.896, 0.998), while ($\text{maxDiff}_{\text{row}}$, $\text{maxDiff}_{\text{col}}$) = (2616.123, 0.498).

(e) The CW blocking has an attractive monotonicity property, apart from being more efficient than RW blocking. For each fixed $n$ ($\leq 100000$) and $\rho$ in $\{0.001, 0.002,\ldots, 0.999\}$, $\text{ESS}_{\text{col}}$ is nondecreasing in $b$, which is intuitively appealing because a larger $b$ in a sense makes block likelihood closer to full likelihood. Contrary to intuition, this monotonicity is not shared by RW blocking, e.g., if $n = 100$ and $\rho = 0.6$, then $\text{ESS}_{\text{row}}$ equals 24.977, 24.763 and 24.361 for $b = 4$, 5 and 10, respectively.

Because of (a)-(d) above, one can reasonably conjecture that $\text{ESS}_{\text{col}} > \text{ESS}_{\text{row}}$ in general. Though all computational evidence goes in support of this, a proof remains elusive due to the complicated forms of $\text{ESS}_{\text{col}}$ and $\text{ESS}_{\text{row}}$ in Theorem 1, as a result of which the plot of $\text{ESS}_{\text{col}}$ – $\text{ESS}_{\text{row}}$ against $\rho$ often has multiple turning points. However, for all practical purposes, we hope that the aforesaid facts (a)-(d), which quantify the gains via CW blocking, will be as informative as such a general result.

*3.3. On the possibility of still more efficient blocking*

One may wonder if, given $n$, $b$ and $m$, there can be another blocking which is more efficient than CW blocking. In view of the facts (b) and (c) in the last subsection, this appears to be unlikely in general. Extensive simulations in search for such a better alternative, if any, show some promise only for a variant of CW blocking. This variant, called modified column-wise (MCW) blocking, is obtained by (i) arranging $1,\ldots, n$ in the natural order as a $b \times m$ array, (ii) then reversing the even rows of this array, and (iii) finally forming blocks as per the columns of the array reached in (ii). For example if $m = 3$ and $b = 5$, then MCW blocking consists of blocks as given by the columns of

$$\begin{array}{ccc} 1 & 2 & 3 \\ 6 & 5 & 4 \\ 7 & 8 & 9 \\ 12 & 11 & 10 \\ 13 & 14 & 15 \end{array}.$$

The difference between MCW and CW blocking is that the latter goes straight from (i) to (iii) above without the reversal of even rows in (ii). Let $\text{Eff}_{\text{mod}}$ denote the efficiency of MCW blocking, defined as in (6). Then the following hold under the AR(1) model.

If $m = 2$, then $\text{Eff}_{\text{mod}} > \text{Eff}_{\text{col}}$, for $2 \leq b \leq 100$ and $\rho$ in $\{0.1, 0.2,\ldots, 0.9\}$, though $\text{Eff}_{\text{mod}}$ exceeds $\text{Eff}_{\text{col}}$ by less than 1%, for every such $\rho$ whenever $b \geq 6$, and their difference tapers off quickly with further increase in $b$. On the other hand, if $m > 2$, then both $\text{Eff}_{\text{mod}} > \text{Eff}_{\text{col}}$ and $\text{Eff}_{\text{mod}} < \text{Eff}_{\text{col}}$ can happen depending on $\rho$. Indeed, then it is also possible that $\text{Eff}_{\text{mod}} < \text{Eff}_{\text{col}}$, for every $\rho$ in $\{0.1, 0.2,\ldots, 0.9\}$, as happens, e.g., with $(b, m) = (5, 50), (10, 35), (20, 30), (30, 25)$ etc. Thus, MCW blocking has a higher efficiency than CW blocking only in some special situations and there too the improvement is often marginal. At the same time, any such gain comes at the cost of much additional computation because MCW blocking does not share the simplifying features of CW blocking as observed in Remark 1. Due to these reasons, we do not consider MCW blocking any more.

*3.4. AR(1) model: case of block size two*

Before concluding this section, we briefly touch upon the case $b = 2$. By Proposition 2, then $\text{ESS}_{\text{row}} = \text{ESS}_{\text{col}}$, under any stationary correlation model as in (7), including the AR(1) model. A perturbed version of RW blocking turns out to be more efficient than both RW and CW blockings in this case. Let $n = 2m$, $b = 2$, and consider a blocking with $m$ blocks

$$\{u, 2m + 1 - u\}\ (u = 1,\ldots, g) \quad \text{and} \quad \{g + 2u - 1, g + 2u\}\ (u = 1,\ldots, m - g),$$

where the positive integer $g$ satisfies $1 \leq g \leq m - 1$. This is called perturbed row-wise blocking of order $g$, or PRW($g$) blocking, as the last $m - g$ blocks also arise in RW blocking when $g$ is even. For example, if $n = 12$, $b = 2$ and $g = 2$, then the six blocks are $\{1,12\}, \{2,11\}, \{3,4\}, \{5,6\}, \{7,8\}$ and $\{9,10\}$, the last four of which also appear in RW blocking.

Under the AR(1) model, PRW($g$) blocking is seen to perform the best vis-à-vis RW and CW blockings when $g = 2$. With $b = 2$ and efficiency defined as in (6), we find that PRW(2) blocking has higher efficiency than RW and CW blockings for every even $n \leq 1000$ and every $\rho$ in $\{0.1, 0.2,\ldots, 0.9\}$, though the gains are less prominent for larger $n$, because then PRW(2) and RW blockings have a vast majority of blocks in common. On a positive note, for the same reason, the computational burden for PRW(2) blocking is also about the same as that for RW blocking when $n$ is large.

**4. More on one-dimensional correlation models**

*4.1. Two more stationary correlation models*

The results under the AR(1) model, based on closed-form expressions for $\text{ESS}_{\text{row}}$ and $\text{ESS}_{\text{col}}$, suggest a similar outcome also for other one-dimensional stationary correlation models as in (7) even if they do not allow analytical derivation. We consider two such models as given by

I. (linear)   $R = (r_{ij})$, with $r_{ij} = 1 - \rho\,|i - j|\ (i, j = 1,\ldots, n)$ and $0 < \rho \leq 1/(n-1)$.

II. (inverse linear)   $R = (r_{ij})$, with $r_{ij} = 1/(1 + \rho\,|i - j|)\ (i, j = 1,\ldots, n)$ and $\rho > 0$.

Like the AR(1) model, both these are stationary and ensure $r_{ij} \geq 0$, with the value of $r_{ij}$ tapering off with increase in the distance $|i - j|$, as one expects intuitively. The rate of tapering, however, is not the same for these models: it is exponential under the AR(1) model, constant under model I, and inverse quadratic under model II. Note that all $r_{ij}$ are positive under model II. The same holds under model I as well, except for the solitary case of $r_{1n} = r_{n1} = 0$ when $\rho = 1/(n-1)$.

Our findings under models I and II are similar to those under the AR(1) model as reported in Subsection 3.2, though the computations are now less extensive due to lack of analytical results as in Theorem 1. We consider ten values of $\rho$ under either model I or model II, namely, $(n-1)\rho = 0.1, 0.2,\ldots,1.0$ under I, and $\rho = 0.2, 0.4,\ldots,2.0$ under II. Accordingly, given $n$, $b$ and $m$, under any of these models, now we write $\text{minEff}_{\text{row}}$ and $\text{minEff}_{\text{col}}$ to denote the smallest possible $\text{Eff}_{\text{row}}$ and $\text{Eff}_{\text{col}}$, respectively, over the corresponding ten values $\rho$. As indicated below, CW blocking again has a clear edge over RW blocking:

(a) Under both models I and II, for every $n \leq 1000$, $3 \leq b \leq n/2$, i.e., $m \geq 2$, and every $\rho$ as stated above, $\text{ESS}_{\text{col}} > \text{ESS}_{\text{row}}$, and hence $\text{Eff}_{\text{col}} > \text{Eff}_{\text{row}}$. The gains can be quite substantial under model I, e.g., if $n = 900$, $(b, m) = (30, 30)$, then under this model, the pair $(\text{Eff}_{\text{row}}, \text{Eff}_{\text{col}})$ equals $(0.923, 0.995)$, $(0.875, 0.991)$, $(0.819, 0.986)$ and $(0.751, 0.979)$ for $(n-1)\rho = 0.4, 0.6, 0.8$ and $1.0$, respectively.

(b) For $n \leq 1000$, the following hold. Under model I, $\text{minEff}_{\text{col}}$ exceeds 0.9, 0.95 and 0.98 whenever $b \geq 6$, 13 and 32, respectively, and $\text{minEff}_{\text{row}}$ can be much smaller even for relatively large $b$. On the other hand, under model II, $\text{minEff}_{\text{col}}$ exceeds 0.95 and 0.98 whenever $b \geq 3$ and 11, respectively, and compared to model I, $\text{minEff}_{\text{row}}$ comes closer to $\text{minEff}_{\text{col}}$. For instance, with $(b, m) = (20, 10)$ and $(30, 30)$, the pair $(\text{minEff}_{\text{row}}, \text{minEff}_{\text{col}})$ equals $(0.753, 0.972)$ and $(0.751, 0.979)$, respectively, under model I, and $(0.946, 0.988)$ and $(0.960, 0.991)$, respectively, under model II.

(c) Additional computations for $n > 1000$ yield results very similar to the above, e.g., if $(b, m) = (60, 40)$, then the pair $(\text{minEff}_{\text{row}}, \text{minEff}_{\text{col}})$ equals $(0.750, 0.989)$ under model I, and $(0.964, 0.994)$ under model II.

(d) Under both models I and II, for each fixed $n$ ($\leq 500$) and $\rho$ as stated above, $\text{ESS}_{\text{col}}$ is nondecreasing in $b$, i.e., the monotonicity property of CW blocking continues to hold under models I and II. Our computations suggest that RW blocking shares this property only under model I but not under model II.

Turning to the case of block size two, as in Subsection 3.4, we find that PRW(2) blocking improves upon the common efficiency of RW and CW blockings, under models I and II as well. Under both models, PRW($g$) blockings for $g > 2$ can lead to further improvement in efficiency. Because such gains are marginal and come at the expense of increased computational burden, we omit the details.

*4.2. Non-equispaced AR(1) model*

Stationary correlation models as in (7), such as the AR(1) model or models I and II in the last subsection implicitly assume that the *n* spatial points are equispaced over $\mathcal{R}$. Such equal spacing is common in practice and, at least for the AR(1) model, is known to be optimal, in the sense of maximizing ESS, among all spacings over a fixed interval in $\mathcal{R}$; see Kiselák and Stehlík (2008).

The spatial points can, however, be non-equispaced in observational studies when they are not chosen by design. To address this issue, we now consider a more general AR(1) model. With *n* spatial points $s_1,\ldots, s_n$, satisfying $s_1 < \ldots < s_n$, it is given by $R = (r_{ij})$, where

$$r_{ij} = \rho^{|s_i - s_j|}, \qquad i, j = 1,\ldots, n, \tag{14}$$

and $0 \le \rho < 1$. Reassuringly, under (14), CW blocking continues to have higher efficiency than RW blocking when the spatial points do not deviate too much from being equispaced. For brevity, only two of many illustrative situations are presented below. Here $t = (s_2 - s_1, s_3 - s_2,\ldots, s_n - s_{n-1})$ and $\bar{t}$ is the mean of the elements of *t*.

(i) $n = 200$, $(b, m) = (20, 10)$, $t = (t_{11}1_{30}^T, t_{12}1_{129}^T, 1_{40}^T)$, with $t_{11} = 0.8$, $t_{12} = 0.9$;

(ii) $n = 384$, $(b, m) = (16, 24)$, $t = (t_{21}1_{100}^T, t_{22}1_{80}^T, 1_{23}^T, t_{22}1_{80}^T, t_{21}1_{100}^T)$, with $t_{21} = 1.2$, $t_{22} = 1.1$.

In (i), *t* is asymmetric about its centre and $\bar{t} < 1$, whereas in (ii), *t* is symmetric about its centre and $\bar{t} > 1$. However, under both (i) and (ii), $\text{Eff}_{\text{col}}$ exceeds $\text{Eff}_{\text{row}}$ for every $\rho$ in $\{0.1, 0.2,\ldots, 0.9\}$, their difference being more prominent, as in the equispaced case, when $\rho$ is relatively large, e.g., for $\rho = 0.6, 0.7, 0.8, 0.9$, the pair ($\text{Eff}_{\text{row}}$, $\text{Eff}_{\text{col}}$) equals (0.948, 0.991), (0.930, 0.988), (0.915, 0.985), (0.914, 0.981) under (i), and (0.956, 0.995), (0.938, 0.993), (0.924, 0.989), (0.931, 0.982) under (ii).

*4.3 Blocks of unequal size*

So far, we have focused on blocking of $n = mb$ spatial points into *m* blocks each of the same size *b*, where $1 < b < n$. This is infeasible if *n* is a prime. Even otherwise, blocks of equal size may become unattractive due to the structure of *n*, e.g., if $n = 2q$, where *q* is a large odd prime, then *b* has to be either 2 or *q*, which may be considered too small or too large. From this perspective, we now consider the situation where *n* is not necessarily an integral multiple of the number of blocks, *m*, and the block sizes are as nearly equal as possible. Writing *b* for the largest integer in $n/m$ and $f = n - mb$, then there are *f* blocks, each of size $b + 1$, and $m - f$ blocks, each of size *b*.

As a generalization of (10), now RW blocking partitions $\{1,\ldots, n\}$ into the *m* blocks

$$\{(u-1)(b+1) + j: j = 1,\ldots, b+1\}, \quad u = 1,\ldots, f, \quad \text{and} \quad \{f + (u-1)b + j: j = 1,\ldots, b\}, \quad u = f+1,\ldots, m.$$

Similarly, generalizing (11), CW blocking now partitions $\{1,\ldots, n\}$ into the *m* blocks

$\{u + (j-1)m: j=1,\ldots, b+1\}$, $u = 1,\ldots, f$, and $\{u + (j-1)m: j=1,\ldots, b\}$, $u = f+1,\ldots, m$.

For example, if $n = 17$, $m = 3$, then $b = 5$, $f = 2$, and the RW blocks are given by the rows of

$$\begin{array}{cccccc} 1 & 2 & 3 & 4 & 5 & 6 \\ 7 & 8 & 9 & 10 & 11 & 12 \\ 13 & 14 & 15 & 16 & 17 \end{array}$$

while the CW blocks are given by the columns of

$$\begin{array}{ccc} 1 & 2 & 3 \\ 4 & 5 & 6 \\ 7 & 8 & 9 \\ 10 & 11 & 12 \\ 13 & 14 & 15 \\ 16 & 17 \end{array}.$$

We compare RW and CW blockings, as indicated above, under three stationary correlation models, namely, the original AR(1) model, $r_{ij} = \rho^{|i-j|}$, with $\rho$ in $\{0.1, 0.2,\ldots, 0.9\}$, and models I and II of Subsection 4.1, each with ten values of $\rho$ as mentioned there. Under such stationary models, both blockings retain to some extent their simplifying features as noted in Remark 1. Their comparison over a range of $n$ and $m$, however, requires more effort because the possibilities for $m$ are now much more numerous than before as $n/m$ need not be an integer. Hence, we consider $n \leq 200$ and $2 \leq m < n/2$, so that each block has size at least two and not all blocks have size two. Our findings, summarized below for this range of $n$ and $m$, greatly resemble their counterparts for blocks of equal size.

(a) Under all three models and for every $\rho$ as stated above, $\text{Eff}_{\text{col}} > \text{Eff}_{\text{row}}$, whenever each block size is at least four. Moreover, $\text{Eff}_{\text{col}} > 0.95$ for every such $\rho$ if each block size is at least 9, 13 or 3, under the AR(1) model, model I or model II, respectively.

(b) Under model I, for every $\rho$ as stated above, $\text{Eff}_{\text{col}}$ always exceeds $\text{Eff}_{\text{row}}$.

(c) Under the AR(1) model or model II, even if $\text{Eff}_{\text{col}} < \text{Eff}_{\text{row}}$, for any $\rho$ in some cases that involve blocks of size two or three, it always holds that $\text{Eff}_{\text{col}} > \text{Eff}_{\text{row}} - 0.011$, so that the shortfall is never significant. At any rate, typically, blocks of size two or three are not of much interest.

(d) In contrast to (c) above, over the aforesaid range of $n$ and $m$ and also beyond it, there are many situations where $\text{Eff}_{\text{col}}$ is appreciably larger $\text{Eff}_{\text{row}}$, especially under the AR(1) model and model I.

As an illustration of the point in (d) above, let $(n, m) = (890, 30)$. Then the pair $(\text{Eff}_{\text{row}}, \text{Eff}_{\text{col}})$ equals $(0.961, 0.999)$, $(0.941, 0.998)$, $(0.918, 0.996)$, $(0.913, 0.992)$ for $\rho = 0.6, 0.7, 0.8, 0.9$, respectively, under the AR(1) model, and $(0.924, 0.995)$, $(0.876, 0.991)$, $(0.819, 0.986)$, $(0.752, 0.979)$ for $(n-1)\rho = 0.4, 0.6, 0.8$ and $1.0$, respectively, under model I. These figures are almost identical to the ones seen earlier for the equal block size case $n = 900$, $(b, m) = (30, 30)$. Similar examples abound.

Matching our intuition, they suggest that the figures for a close-by equal block size case, where computations are less demanding, can well indicate the outcome when $n/m$ is not an integer.

## 5. Two-dimensional correlation models

We now turn to two-dimensional correlation models. The discussion here can, in principle, be extended to the multi-dimensional situation as well at the expense of heavier notation.

Let $R$ be a positive definite correlation matrix of order $n_1 n_2$ ($n_1, n_2 \geq 2$), with rows and columns indexed by ordered pairs $i_1 i_2$ ($i_k = 1, \ldots, n_k$; $k = 1, 2$) representing coordinates of spatial points on a plane. Let $I$ be the set of these $n_1 n_2$ pairs, and suppose $n_k = m_k b_k$ ($k = 1, 2$), where $m_1, b_1, m_2, b_2$ are positive integers. One can extend the concepts of RW and CW blocking to this two-dimensional setup using Cartesian products of sets, in such a manner that they both partition $I$ into $m = m_1 m_2$ blocks, each of size $b = b_1 b_2$. The blocks in RW blocking are

$$B_{u_1 u_2}^{\text{row}} = B_{u_1}^{(1)\text{row}} \times B_{u_2}^{(2)\text{row}} \qquad (u_1 = 1, \ldots, m_1; u_2 = 1, \ldots, m_2), \qquad (15)$$

while those in CW blocking are

$$B_{u_1 u_2}^{\text{col}} = B_{u_1}^{(1)\text{col}} \times B_{u_2}^{(2)\text{col}} \qquad (u_1 = 1, \ldots, m_1; u_2 = 1, \ldots, m_2), \qquad (16)$$

where $\times$ stands for Cartesian product, and for $u_k = 1, \ldots, m_k$ ($k = 1, 2$),

$$B_{u_k}^{(k)\text{row}} = \{ (u_k - 1) b_k + j : j = 1, \ldots, b_k \}, \qquad B_{u_k}^{(k)\text{col}} = \{ u_k + (j-1) m_k : j = 1, \ldots, b_k \}.$$

In view of (10) and (11), for $k = 1, 2$,

$$B^{(k)\text{row}} = \{ B_{u_k}^{(k)\text{row}} : u_k = 1, \ldots, m_k \} \quad \text{and} \quad B^{(k)\text{col}} = \{ B_{u_k}^{(k)\text{col}} : u_k = 1, \ldots, m_k \}, \qquad (17)$$

are RW and CW blockings, respectively, for the set $\{1, \ldots, n_k\}$ in one dimension.

As a simple example, if $n_1 = 8$, $n_2 = 6$, $m_1 = 2$, $b_1 = 4$, $m_2 = 2, b_2 = 3$, then

$$B_1^{(1)\text{row}} = \{1, 2, 3, 4\}, \; B_2^{(1)\text{row}} = \{5, 6, 7, 8\}, \qquad B_1^{(2)\text{row}} = \{1, 2, 3\}, \; B_2^{(2)\text{row}} = \{4, 5, 6\},$$

$$B_1^{(1)\text{col}} = \{1, 3, 5, 7\}, \; B_2^{(1)\text{col}} = \{2, 4, 6, 8\}, \qquad B_1^{(2)\text{col}} = \{1, 3, 5\}, \; B_2^{(2)\text{col}} = \{2, 4, 6\}.$$

Hence by (15), the RW blocking in two dimensions involves the blocks

$\{11, 12, 13, 21, 22, 23, 31, 32, 33, 41, 42, 43\}, \qquad \{14, 15, 16, 24, 25, 26, 34, 35, 36, 44, 35, 46\},$

$\{51, 52, 53, 61, 62, 63, 71, 72, 73, 81, 82, 83\}, \qquad \{54, 55, 56, 64, 65, 66, 74, 75, 76, 84, 85, 86\},$

and by (16), the CW blocking in two dimensions involves the blocks

$\{11, 13, 15, 31, 33, 35, 51, 53, 55, 71, 73, 75\}, \qquad \{12, 14, 16, 32, 34, 36, 52, 54, 56, 72, 74, 76\},$

$\{21, 23, 25, 41, 43, 45, 61, 63, 65, 81, 83, 85\}, \qquad \{22, 24, 26, 42, 44, 46, 62, 64, 66, 82, 84, 86\}.$

In the spirit of Acosta et al. (2021) (see their Subsection 5.3 on Matérn correlation on a plane), RW blocking keeps the paired indices within each block, and hence the points represented by them, close together. On the other hand, in CW blocking these are much more spread out within each block and, as in the one-dimensional situation, this can lead to appreciable gains in $\text{ESS}_B$. Proposition 4 below, which refers to a Kronecker correlation structure, motivates the ideas. We write $\otimes$ for Kronecker product and continue to denote the effective sample sizes under full likelihood inference and under the RW and CW blockings in (15) and (16) by ESS, $\text{ESS}_{\text{row}}$ and $\text{ESS}_{\text{col}}$, respectively.

**Proposition 4**. *Let $R = R^{(1)} \otimes R^{(2)}$, where $R^{(1)}$ and $R^{(2)}$ are correlation matrices of orders $n_1$ and $n_2$, respectively. Then*

(a) $\text{ESS} = \text{ESS}^{(1)}\text{ESS}^{(2)}$, (b) $\text{ESS}_{\text{row}} = \text{ESS}_{\text{row}}^{(1)}\text{ESS}_{\text{row}}^{(2)}$, (c) $\text{ESS}_{\text{col}} = \text{ESS}_{\text{col}}^{(1)}\text{ESS}_{\text{col}}^{(2)}$,

*where $\text{ESS}^{(k)}, \text{ESS}_{\text{row}}^{(k)}$ and $\text{ESS}_{\text{col}}^{(k)}$ are the effective sample sizes, apropos of $R^{(k)}$, under full likelihood inference and under the one-dimensional blockings $B^{(k)\text{row}}$ and $B^{(k)\text{col}}$, respectively, $k = 1,2$.*

*Proof.* The truth of (a) is evident from (1). Next, (b) follows from (2) and (3) noting, after some reflection, that the submatrices of $R$ induced by the RW blocking in (15) are Kronecker products of the corresponding submatrices of $R^{(1)}$ and $R^{(2)}$ induced by $B^{(1)\text{row}}$ and $B^{(2)\text{row}}$ in (17). Similarly, (c) follows invoking (16) and (17). □

In view of Proposition 4 and the facts noted in Sections 3 and 4, if $R = R^{(1)} \otimes R^{(2)}$, with $R^{(1)}$ and $R^{(2)}$ given by the AR(1) model, model I or model II, then CW blocking in (16) has a distinct advantage over RW blocking in (15). Moreover, if both $R^{(1)}$ and $R^{(2)}$ correspond to the AR(1) model, then analytical formulae $\text{ESS}_{\text{row}}$ and $\text{ESS}_{\text{col}}$ emerge from Proposition 4, together with Theorem 1. One specific instance of such a Kronecker correlation structure is the Matérn model based on the $L_1$-distance, with smoothness parameter 1/2, as given by

$$r(i_1 i_2, j_1 j_2) = \exp(-d_1/\phi) = \rho^{d_1}, \quad \phi > 0, \tag{18}$$

where $\rho = \exp(-1/\phi)$ ($0 < \rho < 1$), $r(i_1 i_2, j_1 j_2)$ is the $(i_1 i_2, j_1 j_2)$th element of $R$, and $d_1 = |i_1 - j_1| + |i_2 - j_2|$ is the $L_1$-distance between $i_1 i_2$ and $j_1 j_2$. In this case, $R^{(1)}$ and $R^{(2)}$ are given by the AR(1) model with the same $\rho$.

If $R$ has a Kronecker structure, then by (6) and Proposition 4, the efficiency of RW blocking in (15) equals the product of the efficiencies of the corresponding one-dimensional blockings. The same happens with CW blocking in (16) as well. Thus, the efficiencies in two dimensions are typically smaller than their counterparts in one dimension. This phenomenon, which conforms to the curse of dimensionality, holds also when $R$ does not have a Kronecker structure; see, e.g., Table 1 below. At any rate, as indicated in the previous paragraph and confirmed by the computations reported below, CW blocking continues to outperform RW blocking with regard to efficiency in two dimensions.

Turning to models without a Kronecker correlation structure, we consider Matérn models based on the $L_2$-distance, with smoothness parameter 1/2 and 3/2, as given, respectively, by

$$r(i_1 i_2, j_1 j_2) = \exp(-d_2/\phi) = \rho^{d_2}, \qquad \phi > 0, \tag{19}$$

$$r(i_1 i_2, j_1 j_2) = \{1 + (d_2/\phi)\}\exp(-d_2/\phi) = \{1 - d_2 \log \rho\}\rho^{d_2}, \qquad \phi > 0, \tag{20}$$

where $d_2 = \{(i_1 - j_1)^2 + (i_2 - j_2)^2\}^{1/2}$ and, as in (18), $\rho = \exp(-1/\phi)$ ($0 < \rho < 1$). Along the lines of Remark 1, under these models, CW blocking shares the features of RW blocking that facilitate computation, e.g., by (15), the submatrix of $R$, given by rows $i_1 i_2 \in B^{\text{row}}_{u_1 u_2}$ and columns $j_1 j_2 \in B^{\text{row}}_{v_1 v_2}$, depends on $u_1 u_2$ and $v_1 v_2$ only through $u_1 - v_1$ and $u_2 - v_2$, and by (16), the same happens for CW blocking as well.

Under all three correlation models (18)-(20), we find that $\text{ESS}_{\text{col}} > \text{ESS}_{\text{row}}$, and hence $\text{Eff}_{\text{col}} > \text{Eff}_{\text{row}}$, for every $n \leq 500$ with $b_1, b_2 \geq 3$ and $m_1, m_2 \geq 2$, and every $\rho$ in $\{0.1, 0.2,\ldots, 0.9\}$. The same happens also for $n > 500$ in each of the many situations that we studied. Some examples appear in Table 1. In all these, whether $n \leq 500$ or not, $\text{Eff}_{\text{col}}$ exceeds $\text{Eff}_{\text{row}}$ for every $\rho$ in $\{0.1, 0.2,\ldots, 0.9\}$, but the pair ($\text{Eff}_{\text{row}}$, $\text{Eff}_{\text{col}}$) is shown only for $\rho = 0.6, 0.7, 0.8$ and $0.9$, where they differ more substantially under all three models. Indeed, under the Matérn model (20), based on the $L_2$-distance and with smoothness parameter 3/2, which is attractive due to its smoothness properties, $\text{Eff}_{\text{col}}$ can be appreciably larger than $\text{Eff}_{\text{row}}$ even for smaller $\rho$, e.g., for the case $n = 2400$, $(b_1, b_2, m_1, m_2) = (5, 6, 8, 10)$ in Table 1, the pair ($\text{Eff}_{\text{row}}$, $\text{Eff}_{\text{col}}$) equals (0.890, 0.971), (0.850, 0.950), (0.827, 0.924) and (0.818, 0.895) under this model, for $\rho = 0.2, 0.3, 0.4$ and $0.5$, respectively.

Table 1. *The pair* ($\text{Eff}_{\text{row}}$, $\text{Eff}_{\text{col}}$) *in some examples*

| $n$ | $(b_1, b_2, m_1, m_2)$ | Model | $\rho$ | | | |
|---|---|---|---|---|---|---|
| | | | 0.6 | 0.7 | 0.8 | 0.9 |
| 216 | (6, 4, 3, 3) | (18) | (0.888, 0.943) | (0.858, 0.934) | (0.834, 0.931) | (0.841, 0.943) |
| | | (19) | (0.841, 0.905) | (0.810, 0.898) | (0.787, 0.902) | (0.803, 0.925) |
| | | (20) | (0.750, 0.845) | (0.730, 0.846) | (0.729, 0.863) | (0.781, 0.908) |
| 1680 | (8, 6, 7, 5) | (18) | (0.895, 0.959) | (0.870, 0.943) | (0.843, 0.928) | (0.804, 0.923) |
| | | (19) | (0.862, 0.924) | (0.839, 0.902) | (0.804, 0.888) | (0.751, 0.892) |
| | | (20) | (0.798, 0.867) | (0.775, 0.846) | (0.725, 0.836) | (0.666, 0.858) |
| 2400 | (5, 8, 6, 10) | (18) | (0.899, 0.960) | (0.878, 0.943) | (0.854, 0.923) | (0.817, 0.912) |
| | | (19) | (0.870, 0.926) | (0.850, 0.899) | (0.817, 0.876) | (0.763, 0.874) |
| | | (20) | (0.810, 0.865) | (0.788, 0.837) | (0.738, 0.820) | (0.670, 0.835) |

We finally demonstrate how CW blocking entails improved efficiency in a big data setup. This example relates to the forest dataset considered in Acosta et al. (2021, Section 7). Let $n = 21026304$, $(b_1, b_2, m_1, m_2) = (54, 36, 104, 104)$. Here $n$ is too large to allow computation of ESS under models (19) and (20) which do not have a Kronecker correlation structure. However, invoking the simplifying features of CW and RW blockings in these models as noted earlier, one can obtain $\text{ESS}_{\text{col}}$ and $\text{ESS}_{\text{row}}$, and hence the percentage gain in efficiency via CW blocking over RW blocking as given by $100 \times (\text{ESS}_{\text{col}} - \text{ESS}_{\text{row}})/\text{ESS}_{\text{row}}$. Table 2 shows these percentage gains under models (18)-(20) for $\rho$ in $\{0.1, 0.2,\ldots, 0.9\}$. Again, under all three models, these gains are impressive for relatively large $\rho$, and under model (20), they remain so even for smaller $\rho$.

Table 2. *Percentage gains in efficiency via CW blocking over RW blocking for $n = 21026304$ and $(b_1, b_2, m_1, m_2) = (54, 36, 104, 104)$.*

| Model | $\rho$ | | | | | | | | |
|---|---|---|---|---|---|---|---|---|---|
| | 0.1 | 0.2 | 0.3 | 0.4 | 0.5 | 0.6 | 0.7 | 0.8 | 0.9 |
| (18) | 0.10 | 0.42 | 1.05 | 2.10 | 3.78 | 6.38 | 10.37 | 16.30 | 21.32 |
| (19) | 0.23 | 1.03 | 2.50 | 4.73 | 7.83 | 11.92 | 16.94 | 21.68 | 18.60 |
| (20) | 1.63 | 4.57 | 8.36 | 12.82 | 17.64 | 22.27 | 25.39 | 23.25 | 8.53 |

## 6. Concluding Remarks

In this paper, we explored the impact of the choice of blocks on block likelihood inference for Gaussian random fields with a constant mean. Under a wide variety of correlation models in one and two dimensions, CW blocking was found to be considerably more efficient than RW blocking while retaining the computational simplicity of the latter. Several open issues emerge from the current work.

Under the AR(1) model, all computational evidence based on Theorem 1 suggest that $\text{ESS}_{\text{col}}$ is always greater than $\text{ESS}_{\text{row}}$. A formal proof of this is as yet intractable, but will be of theoretical interest. Furthermore, for any of the one- or two-dimensional correlation models studied here, it will be worthwhile to know if there exists a blocking that shares the simplifying features of CW blocking but enjoys still higher efficiency under wide generality. Again, our findings indicate that no such blocking exists, but any sufficiently general result in this direction will be welcome.

In addition, the points discussed by Acosta et al. (2021, Section 8) deserve attention also in the context of the present paper. The foremost of these concerns block likelihood inference in spatial regression models (Acosta and Vallejos, 2018). As a first attempt towards understanding the role played there by the choice of blocks, we considered one-dimensional stationary correlation models in the special case of a single predictor and no intercept. Then $\text{ESS}_B$ and $\text{Eff}_B$ can be defined along the lines of (2) and (6) replacing the vectors of ones in (1) or (3) by the predictor vector or subvectors thereof, and our initial findings again go in favour of CW blocking. To cite just one of many such examples, with a single predictor vector $z = (z_1,\ldots,z_n)^T$ which plays the role of $1_n$ in Section 2, if $n =$ 900, $z_i \propto i^2$ ($i = 1,\ldots, n$) and we have $m = 30$ blocks each of size $b = 30$, then under the AR(1) model, $\text{ESS}_{\text{col}} > \text{ESS}_{\text{row}}$ for every $\rho$ in $\{0.1,\ldots, 0.9\}$ and the pair ($\text{Eff}_{\text{row}}, \text{Eff}_{\text{col}}$) equals (0.962, 0.997), (0.943, 0.995), (0.919, 0.990), (0.907, 0.979) for $\rho = 0.6, 0.7, 0.8$ and 0.9, respectively. These figures are akin to the ones seen in Subsection 3.2 for $n = 900$, $(b, m) = (30, 30)$, in the setup of a constant mean. The outcome remains similar for models I and II of Section 4, and the gains through CW blocking turn out to be even more significant under model I.

The picture mentioned in the last paragraph, though promising, is of course quite incomplete and, more generally, much will depend on the predictors. Following Acosta and Vallejos (2018), with multiple predictors as well as an intercept, ESS can be defined as the trace of a matrix valued version of (1) after scaling each predictor properly. Similar considerations should also apply to $\text{EES}_B$. Despite these new features, the basic idea behind CW blocking, i.e., spreading out the spatial points within each block in some sense, may continue to be useful.

Another issue mentioned by Acosta et al. (2021) that remains relevant here too concerns the case of irregularly spaced spatial data. It will be of interest to examine how the strategy of spreading out the points within each block can work there retaining, at the same time, the computational advantage of CW blocking as seen in this paper. Further directions for future research include study of alternative sample size reduction techniques as well as extension to non-Gaussian random fields in the presence of asymmetry or heavy tails; see e.g., Bevilacqua and Gaetan (2015), Sun et al. (2018) and Xu and Genton (2017). However, as noted by Acosta et al. (2021), obtaining a convenient expression for the Godambe information is likely to be a challenge in these situations

We conclude with the hope that the present endeavour will generate interest in the above and related problems.

**Appendix**

*Proof of Theorem* 1.

(a) First consider $\text{ESS}_{\text{row}}$. For $u = 1,\ldots, m$, by (10), $R_{uu}$ is $b \times b$, with $(i, j)$th element $\rho^{|i-j|}$ ($i, j = 1, \ldots, b$), so that by Proposition 1(a),

$$R_{uu}^{-1} 1_b = y_b(\rho), \tag{A.1}$$

and hence by (3) and (8),

$$\Sigma_{u=1}^{m} \lambda_{uu} = \Sigma_{u=1}^{m} 1_b^{\text{T}} R_{uu}^{-1} 1_b = m 1_b^{\text{T}} y_b(\rho) = \{n(1-\rho) + 2m\rho\}/(1+\rho). \tag{A.2}$$

Again, by (10), for $u < v$, $R_{uv}$ is $b \times b$, with $(i, j)$th element $\rho^{(v-u)b+j-i}$ ($i, j = 1,\ldots, b$), i.e.,

$$R_{uv} = \rho^{(v-u)b} \xi_1 \xi_2^{\text{T}}, \tag{A.3}$$

where $\xi_1 = (1, \rho^{-1},..., \rho^{-(b-1)})^T$ and $\xi_2 = (1, \rho,..., \rho^{b-1})^T$. Because by (8), $\xi_1^T y_b(\rho) = \rho^{-(b-1)}$ and $\xi_2^T y_b(\rho) = 1$, from (3), (A.1) and (A.3), one obtains $\lambda_{uv} = \rho^{(v-u)b-(b-1)}$, for $u < v$. As a result, for $u > v$, the symmetry of $R$ yields $\lambda_{uv} = \rho^{(u-v)b-(b-1)}$. Thus, $\lambda_{uv} = \rho^{|u-v|b-(b-1)}$, for every $u \neq v$. So,

$$\Sigma_{u=1}^m \Sigma_{v(\neq u)=1}^m \lambda_{uv} = \rho^{-(b-1)} \Sigma_{u=1}^m \Sigma_{v(\neq u)=1}^m \rho^{|u-v|b} = \rho^{-(b-1)}(1_m^T \tilde{R} 1_m - m), \quad (A.4)$$

where $\tilde{R}$ is the $m \times m$ correlation matrix under an AR(1) model with $\rho$ changed to $\rho^b$. Since by (13),

$$1_m^T \tilde{R} 1_m = \frac{m(1-\rho^{2b}) - 2\rho^b(1-\rho^{mb})}{(1-\rho^b)^2} = \frac{1}{1-\rho^b}\left(m(1+\rho^b) - \frac{2\rho^b(1-\rho^n)}{1-\rho^b}\right),$$

from (A.4) one gets

$$\Sigma_{u=1}^m \Sigma_{v(\neq u)=1}^m \lambda_{uv} = \frac{2\rho}{1-\rho^b}\left(m - \frac{1-\rho^n}{1-\rho^b}\right). \quad (A.5)$$

Substituting (A.2) and (A.5) in (2), the expression for ESS$_{\text{row}}$ follows.

(b) We next consider ESS$_{\text{col}}$. For $u = 1,..., m$, by (11), $R_{uu}$ is $b \times b$, with $(i, j)$th element $\rho^{|i-j|m}$ ($i, j = 1,..., b$), so that analogously to (A.1) and (A.2) by Proposition 1(a),

$$R_{uu}^{-1} 1_b = y_b(\rho^m), \quad (A.6)$$

and

$$\Sigma_{u=1}^m \lambda_{uu} = \{n(1-\rho^m) + 2m\rho^m\}/(1+\rho^m). \quad (A.7)$$

The $R_{uv}$, however, no longer admit a factorization as in (A.3), and hence their handling now requires more effort. To that effect, observe that by (3), (8) and (A.6), for $u, v = 1,..., m$,

$$\lambda_{uv} = y_b(\rho^m)^T R_{uv} y_b(\rho^m)$$
$$= \{(1-\rho^m)1_b + \rho^m \psi_b\}^T R_{uv}\{(1-\rho^m)1_b + \rho^m \psi_b\}/(1+\rho^m)^2,$$
$$= \{(1-\rho^m)^2 1_b^T R_{uv} 1_b + 2\rho^m(1-\rho^m)\psi_b^T R_{uv} 1_b + \rho^{2m} \psi_b^T R_{uv} \psi_b\}/(1+\rho^m)^2. \quad (A.8)$$

Now, by (13),

$$\Sigma_{u=1}^m \Sigma_{v=1}^m 1_b^T R_{uv} 1_b = 1_n^T R 1_n = \{n(1-\rho^2) - 2\rho(1-\rho^n)\}/(1-\rho)^2. \quad (A.9)$$

Next, recall that $R[i]$ denotes the sum of the elements of the $i$th row of $R$, and from (11) observe that, for $u = 1,..., m$, the vector $\Sigma_{v=1}^m R_{uv} 1_b$ is $b \times 1$, with $j$th element given by $R[u + (j-1)m]$ ($j = 1,..., b$). Hence, by (9) and the definition of $\psi_b$,

$$\psi_b^T (\Sigma_{v=1}^m R_{uv} 1_b) = R[u] + R[u + (b-1)m]$$
$$= \{2(1+\rho) - \rho^u - \rho^{n-u+1} - \rho^{u+(b-1)m} - \rho^{n-u-(b-1)m+1}\}/(1-\rho).$$

Since $n = mb$, summing the above over $u = 1,..., m$, after some simplification,

$$\Sigma_{u=1}^{m}\Sigma_{v=1}^{m}\psi_b^T R_{uv} 1_b = 2\{m(1-\rho^2) - \rho(1+\rho^{n-m})(1-\rho^m)\}/(1-\rho)^2. \tag{A.10}$$

For $u, v = 1,\ldots, m$, from (11), now note that the (1,1)th, (1,$b$)th, ($b$,1)th and ($b$,$b$)th elements of $R_{uv}$ equal $\rho^{|u-v|}$, $\rho^{|u-v-(b-1)m|}$, $\rho^{|u-v+(b-1)m|}$ and $\rho^{|u-v|}$, respectively. Since $v - u + (b - 1)m$ and $u - v + (b - 1)m$ and are both nonnegative as $b \geq 3$ and $|u - v| \leq m - 1$, recalling the definition of $\psi_b$,

$$\Sigma_{u=1}^{m}\Sigma_{v=1}^{m}\psi_b^T R_{uv}\psi_b = \Sigma_{u=1}^{m}\Sigma_{v=1}^{m}(2\rho^{|u-v|} + \rho^{v-u+(b-1)m} + \rho^{u-v+(b-1)m})$$

$$= 2\Sigma_{u=1}^{m}\Sigma_{v=1}^{m}(\rho^{|u-v|} + \rho^{u-v+(b-1)m}\}.$$

$$= 2\{m(1-\rho^2) - 2\rho(1-\rho^m) + \rho^{n-2m+1}(1-\rho^m)^2\}/(1-\rho)^2, \tag{A.11}$$

because

$$\Sigma_{u=1}^{m}\Sigma_{v=1}^{m}\rho^{|u-v|} = \{m(1-\rho^2) - 2\rho(1-\rho^m)\}/(1-\rho)^2,$$

analogously to (13), and

$$\Sigma_{u=1}^{m}\Sigma_{v=1}^{m}\rho^{u-v+(b-1)m} = \rho^{(b-1)m}(\Sigma_{u=1}^{m}\rho^u)(\Sigma_{v=1}^{m}\rho^{-v}) = \rho^{n-2m+1}(1-\rho^m)^2/(1-\rho)^2.$$

Finally, by (A.8)-(A.11), after a long algebra,

$$\Sigma_{u=1}^{m}\Sigma_{v=1}^{m}\lambda_{uv} = \frac{(1-\rho^2)\{(n-2m)(1-\rho^m)^2 + 2m\} - 2\rho(1-\rho^{2m})}{(1+\rho^m)^2(1-\rho)^2}. \tag{A.12}$$

Substituting (A.7) and (A.12) in (2), the expression for $\text{ESS}_{\text{col}}$ follows. □

**Acknowledgement**. This work was supported by a grant from the Science and Engineering Research Board, Government of India.